\documentclass[pra,aps,amsmath,amssymb,showpacs]{revtex4}
\usepackage{graphicx}
\newcommand{\aver}[1]{\left< #1 \right>}
\begin{document}
\title{Level spacing statistics in a randomly-inhomogeneous acoustic waveguide}
\author{D.V. Makarov}
\email{makarov@poi.dvo.ru}
\homepage{http://dynalab.poi.dvo.ru}
\affiliation{Laboratory of Nonlinear Dynamical Systems,\\
V.I. Il'ichev Pacific Oceanological Institute of the Russian Academy of
Sciences,\\
690041 Vladivostok, Russia}
\author{L.E.~Kon'kov}
\affiliation{Laboratory of Nonlinear Dynamical Systems,\\
V.I. Il'ichev Pacific Oceanological Institute of the Russian Academy of
Sciences,\\
690041 Vladivostok, Russia}
\author{M.Yu.~Uleysky}
\affiliation{Laboratory of Nonlinear Dynamical Systems,\\
V.I. Il'ichev Pacific Oceanological Institute of the Russian Academy of
Sciences,\\
690041 Vladivostok, Russia}
\begin{abstract}
Dynamics of a randomly-perturbed quantum system with
3/2-degrees of freedom is considered.
We introduce a transfer operator being the quantum analogue of the specific
Poincar\'e map. This map was proposed in
(Makarov, Uleysky, J.~Phys.~A: Math.~Gen., 2006)
for exploring domains of finite-time stability, which survive
under random excitation.
Our attention is concentrated on
level spacing distribution of the transfer operator,
averaged over ensemble of realizations.
The problem of sound propagation in an oceanic waveguide
is considered as an example.
For the acoustic frequency of 200 Hz,
level spacing distribution undergoes the crossover from
Poissonian to Wigner-like statistics with increasing time, as it is consistent with
classical predictions via the specific Poincar\'e map.
For the frequency of 600 Hz,
the level spacing statistics becomes non-universal due to
large number of nearly-degenerate levels whose contribution grows with time.
Occurrence of nearly-degenerate levels is attributed to
bifurcations of classical periodic orbits, caused by
fast spatial oscillations of the random perturbation.
\end{abstract}
\maketitle

Manifestations of classical chaos in quantum motion
are one of the most intriguing fields in the quantum theory.
It is recognized that quantum behavior qualitatively depends
on whether the classical counterpart is regular or chaotic.
Regular classical motion typically corresponds to
laminar and coherent quantum evolution.
In contrast,
chaos in the classical limit gives rise to diffusive wavepacket spreading
\cite{Izrailev,Kolovsky}, fast decoherence
\cite{Peres}, 
and irregular nodal patterns of eigenfunctions \cite{Smilansky}.
In quantum spectra, classical chaos infers
repulsion of the nearest energy levels
and Wigner-like statistics of level spacings,
as opposed to Poissonian statistics in the integrable case
\cite{Stockman}.
Thus, level spacing distribution is a good indicator
of chaotic manifestations into quantum motion.
Hamiltonian systems usually have mixed
phase space with the coexistence of regular and chaotic domains.
In the semiclassical limit, level sequences contributed
from regular and chaotic domains
are independent, and spacing distribution obeys the Berry-Robnik law \cite{BR},
deviating from it
as quantum corrections grow \cite{Prosen}.

If the quantum system is subjected to a random perturbation,
it is reasonable to expect purely chaotic diffusive behavior
due to the absence of stable domains \cite{Kolovsky}.
However, if we deal with finite-time evolution,
there can be phase space domains where Lyapunov stability and
coherent motion still survive.
They can be found by
computing distribution of finite-time Lyapunov exponents in phase space,
or finding eigenfunctions of the classical transfer operator \cite{Froyland}.
In this Letter we consider the method
based on the construction
of the specific Poincar\'e map
for a typical realization of the random force \cite{JPA,PRE73}.
We describe briefly the main scenarios of global chaos
and examine how they are reflected on the quantum level.
In particular,
we introduce the quantum version of the specific Poincar\'e map,
and study how its spectral statistics depends on the length
of the time interval.
It is shown that this dependence can be non-trivial
when classical chaos is caused by cascade of periodic orbit bifurcations.
Sound propagation in an oceanic waveguide is considered as an example.

We begin with the classical level.
Consider the Hamiltonian of a unit-mass particle
\begin{equation}
H=\frac{p^2}{2}+U(q)+\varepsilon V(q)\xi(q,t),
\label{ham01}
\end{equation}
where $p$ and $q$ are momentum and position, respectively,
$U(q)$ is an unperturbed potential,
$\varepsilon\ll 1$, $V(q)$ is some smooth function,
and $\xi(q,t)$ is a zero-mean random function.
Assume that $\xi$ is differentiable, stationary,
and normalized as $\aver\xi^2=1$.
If we consider a single realization of $\xi$,
we can regard it as a deterministic function with a broad temporal spectrum.
According to \cite{JPA,PRE73},
one can seek for those sets in phase space that transform to themselves
without mixing in the course of evolution from $t=t_0$ to $t=t_0+\tau$.
Such sets correspond to trajectories that stable by Lyapunov
within the interval $[t_0:t_0+\tau]$.
They can be found using the map
\begin{equation}
p_{i+1}=p(t_0+\tau |\,p_i,q_i),\quad q_{i+1}=q(t_0+\tau |\,p_i,q_i),
\label{map}
\end{equation}
where $p(t_0+\tau |\,p_i,q_i)$ and $q(t_0+\tau |\,p_i,q_i)$
are the solutions of the Hamiltonian equations
at $t=t_0+\tau$ with initial conditions $p(t_0)=p_i$, $q(t_0)=q_i$.
It means that values of $p$ and $q$ on the $i$-th step of mapping
become the initial conditions for the $i+1$-th step.
After the time shift $t'=t-t_0$,
the specific Poincar\'e map (\ref{map})
becomes equivalent to the usual Poincar\'e
map corresponding to the Hamiltonian
\begin{equation}
\bar H=\frac{p^2}{2}+U(q)+\varepsilon V(q)\tilde\xi(q,t')
\label{ham-period}
\end{equation}
where $\tilde\xi(t')$ is periodic function being a train
of identical pieces of $\xi(t)$
\begin{equation}
\tilde\xi(q,t'+n\tau)=\xi(q,t'),\quad
0\le t'\le\tau ,
\label{xi-sr}
\end{equation}
with $n$ being an integer.
Thus, we replace the original Hamiltonian by a periodically-driven
one. This replacement is justified by the restriction to examine
evolution only within the interval $[t_0:t_0+\tau]$.
The main property of the map (\ref{map}):
{\it each point of a continuous closed invariant set of the map (\ref{map})
corresponds to an initial point of a trajectory
that is stable by Lyapunov
within the interval $[t_0:t_0+\tau]$}.
The map (\ref{map})
provides a sufficient but not necessary criterion of stability.

There can be two different scenarios of chaos onset for the map (\ref{map}),
depending on whether $\tilde\xi$ varies with $q$ slowly or fastly.
For slow variations, transition to global chaos can be described
by means of overlapping of resonances \cite{Chirikov,Zas}.
It is convenient to transform variables $p-q$ to the action-angle
variables $I-\vartheta$ \cite{Zas}.
Resonances are determined by the condition
\begin{equation}
m\tau=nT(I=I_\text{res}),
\label{rescond}
\end{equation}
where $m$ and $n$ are integers,
and $T$ is a period of unperturbed oscillations.
Strength of a resonance
can be quantified by the modulo of the Fourier-amplitude
\begin{equation}
H_{m,n}=\frac{1}{2\pi\tau}\int\limits_{\vartheta=0}^{2\pi}
\int\limits_{t'=0}^{\tau}
V\tilde\xi\,e^{-i(m\vartheta+n\nu t')}
d\vartheta dt',
\label{fourier}
\end{equation}
where $\nu=2\pi/\tau$.
It is natural to expect that $|H_{m,n}|$ decays,
on average, with $m$ and $n$, therefore, only some finite set of
dominant resonances (\ref{rescond}) should be taken into account.
The case of $\tau\ll T$ corresponds to high-frequency driving when
all resonances are weak and can be neglected.
Then the Hamiltonian (\ref{ham-period}) can be reduced to a time-independent
and integrable one using the averaging method \cite{BM}.
In the limit $\tau\gg T$ a distance between neighboring resonances (\ref{rescond})
scales as $\tau^{-1}$, and they eventually overlap completely, i.~e.,
motion becomes fully chaotic according to the Chirikov's criterion.
However, small islands of stability can persist
even on timescales large compared with $T$ \cite{JPA,PRE73}.

Fast oscillations of $\tilde\xi$ with $q$ (and, thus, with $\vartheta$)
lead to a more complicated scenario.
In this case the integration over $\vartheta$
in (\ref{fourier}) can be carried out using the stationary
phase techniques.
The respective stationary phase condition
can be easily derived if we substitute
$\tilde\xi=\Upsilon\exp(i\alpha)+\mathrm{c.~c.}$,
where $\Upsilon$ is a slowly-varying term,
into (\ref{fourier}).
Then neglecting the slow dependencies
of $V$ and $\Upsilon$ on $\vartheta$, we obtain
the desired condition
$\partial\alpha/\partial\vartheta\pm m=0$.
If this condition is fulfilled simultaneously
with vanishing $\partial^2\alpha/\partial\vartheta^2$
(the second derivative of the phase in (\ref{fourier})),
certain resonances (\ref{rescond}) are enhanced singularly,
as well as the corresponding
derivatives $|\partial H_{m,n}/\partial I|$ and
$|\partial^2 H_{m,n}/\partial I^2|$.
These derivatives quantify perturbation-induced variations of the frequency,
and their enhancement can cause violation of nondegeneracy,
i.~e. leads to nonmonotonic dependence of frequency of oscillations on the action.
This phenomenon is closely linked to the occurrence of
nonstationary wavelike distortions of the potential, caused
by the perturbation imposed.
These distortions can scatter
classical trajectories making them chaotic.
As one considers how classical dynamics changes with increasing $\varepsilon$,
the nondegeneracy violation in vicinities of resonances (\ref{rescond})
corresponds to multiplication of fixed points
due to the cascade of pitchfork and saddle-center bifurcations.
As a result, there arise clusters of splitted orbits.
If the bifurcations appear, global chaos can emerge even for low $\tau$.
As $\tau$ increases, the number of orbits undergoing bifurcations grows.
A more detailed description of this scenario
can be found in \cite{Book,EPJ}.

One of the most important properties of the specific
Poincar\'e map is that area of the overall regular domain
weakly depends on a realization of $\tilde\xi$, albeit details
of a phase space pattern are typically different. This circumstance
validates the usage of such a quasideterministic
map for studying randomly-driven
systems.
A more comprehensive analysis of the specific Poincar\'e map
can be found in \cite{Book}.
In \cite{Gan}, the specific Poincar\'e map was used in the presence of
dissipation.

Following the analogy with the specific Poincar\'e map,
one can introduce a quantum transfer operator $\hat F$
which describes evolution of the quantum state
between time instants $t_0$ and $t_0+\tau$ under a perturbation
involving a particular realization of the random function $\xi(t)$
\begin{equation}
 \hat F\Phi(q,t_0+\tau)=\Phi(q,t_0).
\label{operator}
\end{equation}
It is analogous to the Floquet operator corresponding to a classical Hamiltonian (\ref{ham-period}).
One can define eigenstates and eigenvalues of the operator $\hat F$ by means of the equation
\begin{equation}
\hat F\Phi_m(q,t_0+\tau)=e^{-i\epsilon_m/\hbar}\Phi_m(q,t_0),
\label{eigen}
\end{equation}
where quantity $\epsilon_m$ plays the role of quasienergy.
In the semiclassical limit $\hbar\to 0$, quantum behavior described by the operator $\hat F$
is consistent with the classical map (\ref{map}).
So, one should expect some signatures of the transient stable domains
in the spectrum of $\hat F$.
We can introduce ensemble-averaged level spacing distribution
\begin{equation}
\rho(s,\tau)=\lim\limits_{N\to\infty}\frac{1}{N}\sum\limits_{n=1}^{N} P_n(s,\tau),
\label{aver_ps}
\end{equation}
where $s$ is a difference between two neighboring values of quasienergy,
$P_n(s,\tau)$ is a level spacing distribution for $\hat F$ calculated with $n$-th realization
of $\xi(q,t)$. $\rho(s,\tau)$ can serve as a representative characteristics describing quantum evolution
within the temporal interval of a length $\tau$.
Transformation of $\rho(s,\tau)$ with increasing $\tau$ should be attributed
to the crossover from an initial laminar phase to intermittency, and, eventually,
to ergodic diffusive behavior \cite{Kolovsky}.
\begin{figure}[!h]
\centerline{\includegraphics[width=0.95\textwidth]{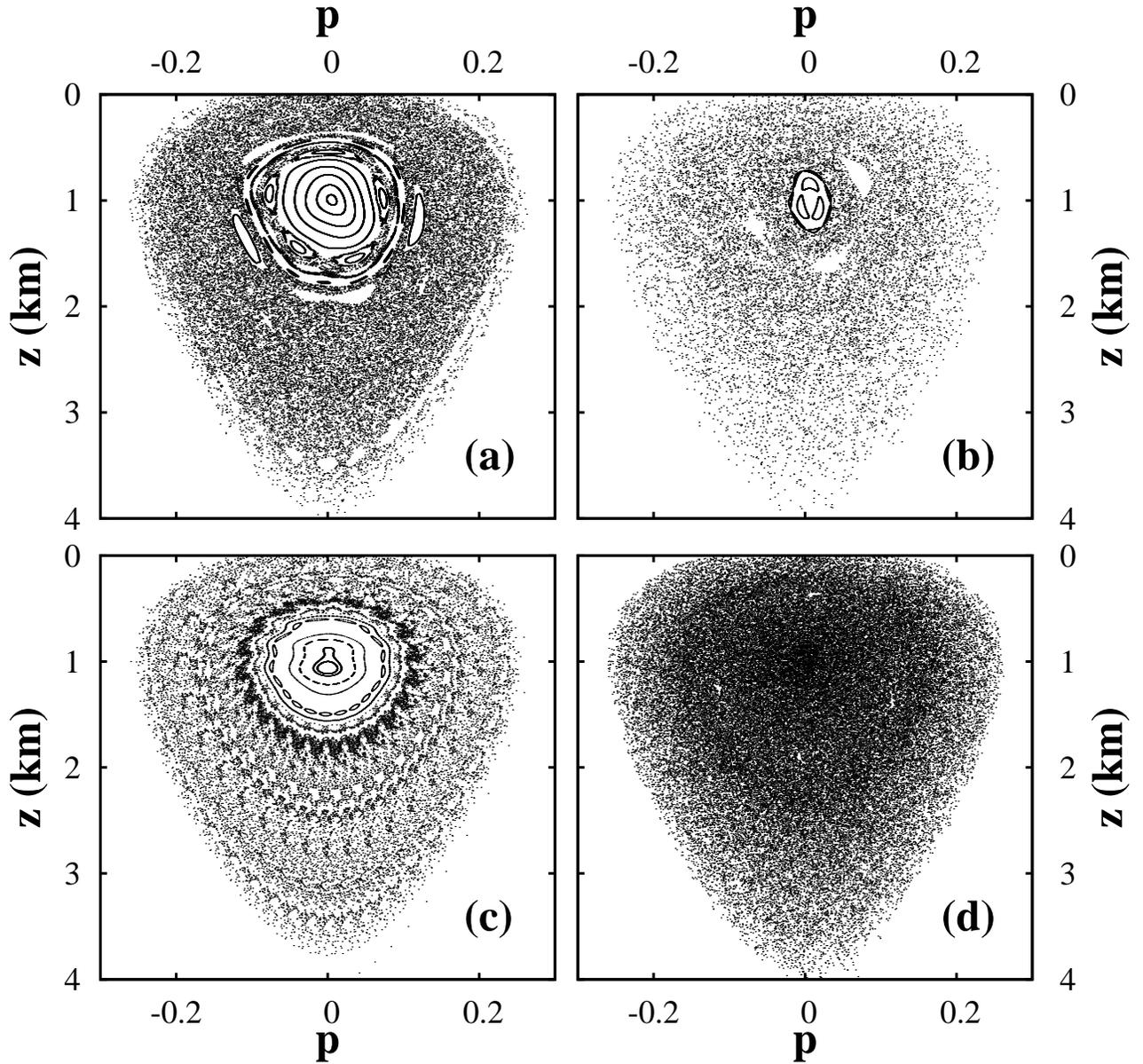}}
\caption{Examples of phase space portraits constructed
via the specific Poincar\'e map.
The values of parameters are: (a) $u=5$, $\tau=100$~km,
(b) $u=5$, $\tau=500$~km, (c) $u=20$, $\tau=10$~km, (d) $u=20$, $\tau=30$~km.}
\label{poinc}
\end{figure}

The example we represent in this paper doesn't relate to quantum mechanics directly but is equivalent
to quantum-mechanical problems in the mathematical sense. That is 
long-range sound propagation
in a randomly-inhomogeneous acoustic waveguide in an ocean.
The waveguide occurs due to the nonmonotonic depth dependence
of the sound speed, with a minimum at some depth,
and maintains sound waves inside water column with weak attenuation.
Wavefield is described by the standard parabolic equation
%
\begin{equation}
\begin{aligned}
&\frac{i}{k_0}\frac{\partial\Phi}{\partial r}=\hat H\Phi,\quad
\hat H=-\frac{1}{2k_0^2}\frac{\partial^2}{\partial z^2}+U(z,r),\\
&U(z,r)=\frac{1}{c_0}\left[\Delta c(z)+\delta c(z,\,r)\right].
\end{aligned}
\label{parabolic}
\end{equation}
Here $k_0=2\pi f/c_0$ is the wavenumber
in the reference medium with $c=c_0$, $f$ is signal frequency.
The parabolic equation formally coincides with
the time-dependent Shr\"odinger equation for a unit-mass quantum particle.
In this analogy, range $r$ plays the role of
the timelike variable, the inverse wavenumber $k_0^{-1}$
serves as the Planck constant, and
$U(z,r)$ is a potential consisted of
an unperturbed part $\Delta c(z)/c_0$ and a small perturbation
$\delta c(z,r)/c_0$ caused by oceanic internal waves.

Ray trajectories obey the Hamiltonian equations
\begin{equation}
\dfrac{dz}{dr}=\dfrac{\partial H}{\partial p},\quad
\dfrac{dp}{dr}=-\dfrac{\partial H}{\partial z},\quad
H=\frac{p^2}{2}+U(z,r),
\label{h}
\end{equation}
where $p=\tan\chi$ is the analog to mechanical momentum,
$\chi$ is a grazing angle of a sound ray.
Perturbation becomes a significant factor
on long ranges, inducing ray chaos \cite{Book,Review}.
We use the following expression for $\delta c$:
\begin{equation}
\delta c(z,r)=\varepsilon c_0\frac{z}{B}e^{-\frac{2z}{B}}
\sin\left[\pi\biggl(ue^{-\frac{z}{B}}+\mu(r)\biggr)\right]\mu(r).
 \label{dc}
\end{equation}
Here $\varepsilon$ is a small constant taken of $0.0014$, $B=1$~km is a thermocline depth,
and $\mu(r)$ is a random function
being a sum of 10000 randomly-phased harmonics
with horizontal wavenumbers $k_r$ distributed
in the range from $2\pi/100$~km$^{-1}$ to
$2\pi/1$~km$^{-1}$, with spectral density decaying as $k_r^{-2}$.
$\mu$ is normalized as $\aver{\mu^2}=1$.
Despite the model (\ref{dc}) is artificial and idealized,
it maintains the main properties of realistic internal-wave fields,
which are responsible for sound scattering,
such as the presence of vertical oscillations with a depth-dependent wavenumber.
Rate of these oscillations depends on the parameter $u$.
We consider two cases: $u=5$ (moderate oscillations) and $u=20$
(fast oscillations).
Unperturbed sound-speed profile is taken in the form \cite{Book}
\begin{equation}
\Delta c(z)=
\dfrac{c_0b^2}{2}\left(e^{-az}-\eta\right)^2,
\quad 0\le z\le h,
\label{prof}
\end{equation}
where $h=4.0$~km is the depth of the ocean bottom,
$a=0.5$~km$^{-1}$, $\eta=0.6065$, $c_0=1480$~m/s.
Ray cycle length (the analogue of the period $T$)
in the unperturbed waveguide varies from 37 km to 55 km.
Fig.~\ref{poinc} represents examples of Poincar\'e plots.
In the case of $u=5$, the Chirikov's scenario prevails.
Area of the regular domain slowly decreases with increasing $\tau$,
and small islands of stability survive for the values of $\tau$ of 
hundreds kilometers.
When $u=20$, the scenario of bifurcation-induced chaos dominates,
and all islands of stability cease completely already for $\tau=30$~km
(see Fig.~\ref{poinc}(d)).
\begin{figure}[!h]
\centerline{\includegraphics[width=0.95\textwidth]{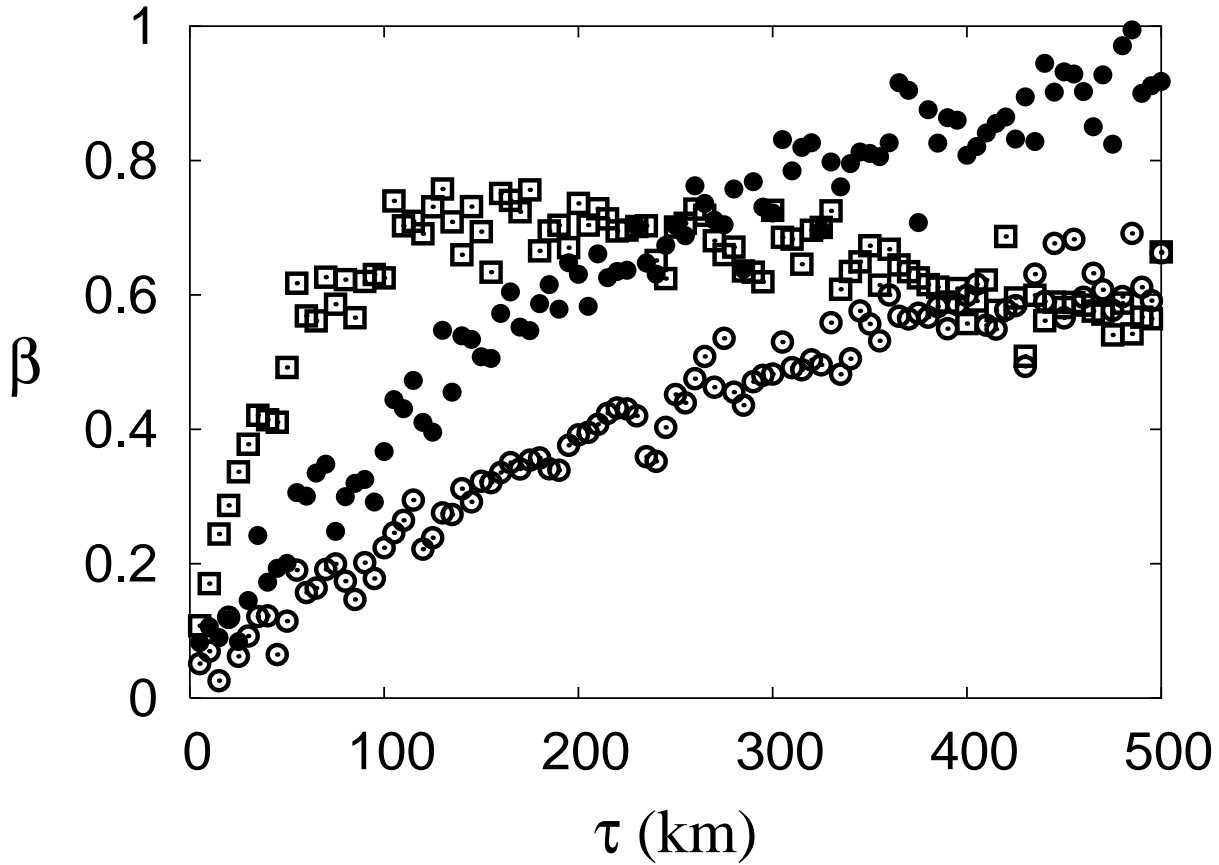}}
\caption{Brody parameter $\beta$ vs $\tau$ for $f=200$~Hz, $u=5$
(empty circles), $f=200$~Hz, $u=20$ (filled circles),
and $f=600$~Hz, $u=20$ (empty squares). } \label{brody}
\end{figure}

In order to examine how peculiarities of classical ray dynamics reveal
themselves with finite wavelengths, we solve numerically
the parabolic equation (\ref{parabolic}) with different
realizations of $\mu$, and then construct the respective operators $\hat F$
in the basis of eigenfunctions of the unperturbed Hamiltonian $\hat H$.
Reflections off the ocean surface $z=0$ are fairly modeled
using the Dirichlet boundary
condition $\Phi=0$.
Here we are mainly interested in nearly-horizontal propagation,
so sound absorption at the bottom can be disregarded,
and the bottom is treated
as the horizontal perfectly-reflecting rigid wall
with the Neumann boundary condition $d\Phi/dz=0$.
\begin{figure}[!t]
\centerline{\includegraphics[width=0.95\textwidth]{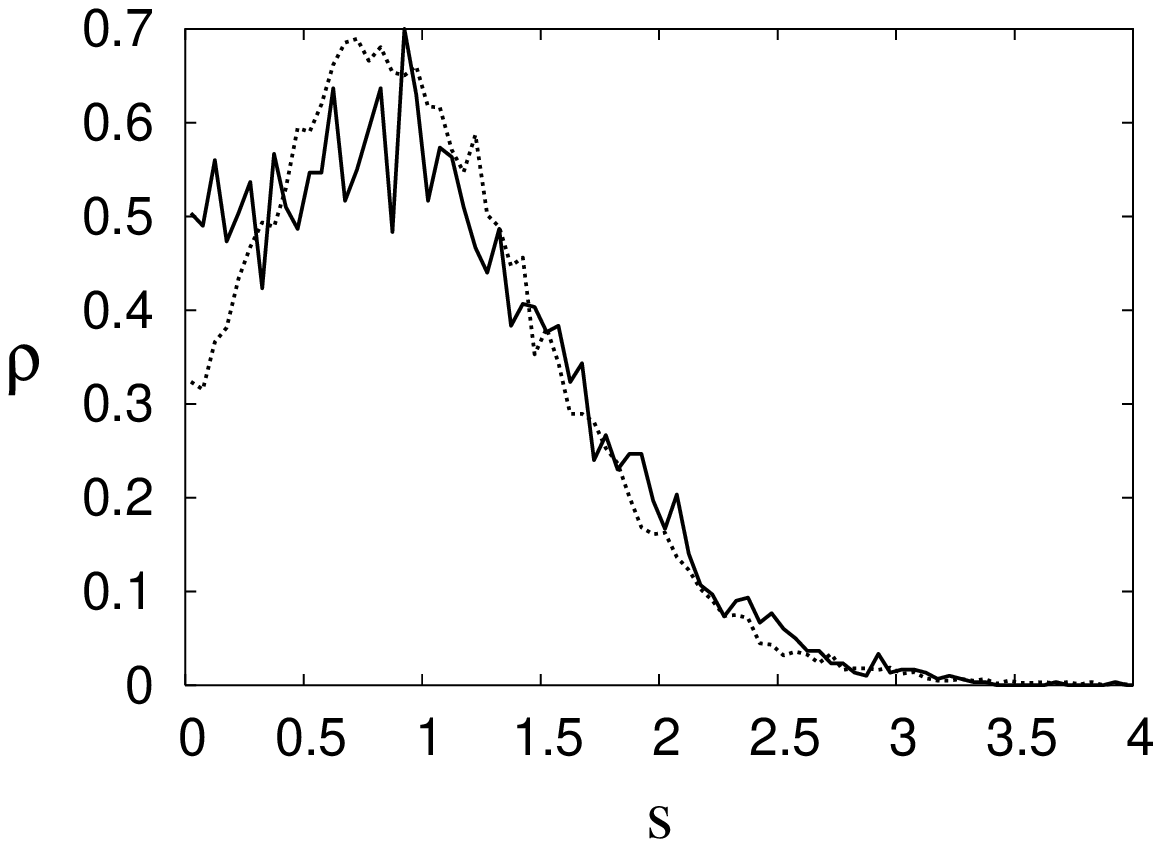}}
\caption{Averaged level spacing distributions computed with $f=600$ Hz,
$u=20$, $\tau=120$~km (dotted) and $\tau=500$~km (solid).}
\label{s}
\end{figure}
We calculated $\rho(s,\tau)$ for $\tau$
ranging from 5 to 500 km.
Number of realizations of $\xi$
decreased with increasing $\tau$ from 1000 to 10 due to
the limitations of computational facilities.
For each realization,
the number of levels taken into account is 200 for $f=200$~Hz,
and 600 for $f=600$~Hz.
The dependence $\rho(s)$ obtained with each $\tau$
is fitted to the Brody distribution
\begin{equation}
 P_\mathrm{B}(s)=(\beta+1)A_{\beta} s^\beta\exp(-A_\beta s^{\beta+1}),
\label{distr}
\end{equation}
where $A_\beta=[\Gamma((\beta+2)/(\beta+1))]^{\beta+1}$, and
$\Gamma$ is the Euler gamma function. The cases $\beta=0$ and
$\beta=1$ correspond to Poissonian and Wigner
distributions, respectively. Figure \ref{brody} shows how
the parameter $\beta$ corresponding to the best fit depends on $\tau$.
For $f=200$~Hz, $\beta(\tau)$ reveals
the crossover from Poissonian to
Wigner-like statistics.
The fluctuations of $\beta$ indicate on the fact that,
according to the Gutzwiller theory \cite{Gutz}, spectrum of the
operator $\hat F$ depends on the spectrum of periodic orbits of
the map $(\ref{map})$, which varies with $\tau$, and these
variations infer some weak nonmonotonicity in the behavior of
$\beta$. The growth is slower for $u=5$ due to the presence of
long-living regular domains (see Fig.~\ref{poinc}(b)). For $u=20$,
$\beta$ increases slower than it is expected by the specific
Poincar\'e map. This can be regarded as a manifestation of
dynamical localization \cite{Izrailev,Stockman}. For $f=600$~Hz
and $u=20$ the behavior is different: $\beta$ increases rapidly
for low $\tau$ and then, for $\tau>150$~km, decreases.
Fig.~\ref{s} shows that $\beta$ decreases due to the growth of the
number of nearly-degenerate levels. This phenomenon can be thought
of as the consequence of large fluctuations of spectral density,
caused by periodic orbit bifurcations \cite{BKP}.
This effect is not seen with
$f=200$~Hz because the majority of bifurcated orbits appear
poorly isolated and don't give individual contributions into
spectrum.
For $f=600$~Hz and $u=5$ the influence of bifurcations is much less
pronounced, therefore, $\beta$ grows until $\tau\simeq 400$~km, and only then
starts decreasing (not shown).

In summary, we calculated level spacing distribution
for the operator which is the quantum analogue of the specific Poincar\'e map and
describes evolution of a randomly-driven quantum system between two
time instants.
If global classical chaos 
emerges solely according to the Chirikov's scenario,
one expects the transition from Poissonian to Wigner-like
statistics with increasing the length of the time interval.
As an example, we considered
guided sound propagation in the ocean,
when fast depth oscillations of the sound-speed perturbation
give rise to cascade of periodic orbit bifurcations.
As a result, there occur large non-universal spectral fluctuations,
and level spacing distribution
for high acoustic frequencies
doesn't exhibit the transition from Poissonian to Wigner-like
statistics, despite of the onset of ergodic chaos.
However, this transition is recovered for smaller acoustic frequencies
due to the wave-based suppression of bifurcation influence.
We suppose that transformation
of $\rho(s)$ with $\tau$ can efficiently describe
quantum decoherence in systems subjected to temporal noise,
when Chirikov's scenario should dominate.

This work was supported by the grants from the President of the Russian Federation (project MK-4324.2009.5),
RFBR--FEB RAS (project 09-05-98608), and the ``Dynasty'' foundation.
Authors are grateful to A.R.~Kolovsky, K.V.~Koshel, S.V.~Prants,
and P.S.~Petrov for helpful discussions and assistance
in the course of this research.

\end{document}